# A Wideband Quasi-Asymmetric Doherty Power Amplifier with a Two-Section Matching-Phase Difference Compensator Network Design Using GaAs Technology

Seyedehmarzieh Rouhani, Ahmad Ghanaatian, Adib Abrishamifar[1], Majid Tayarani

School of Electrical Engineering, Iran University of Science and Technology, Tehran, Iran, Tel: +98(21) 73225727, Fax: +98(21) 73225777

*Abstract-* In this paper, a quasi-asymmetric Doherty power amplifier (PA) is designed without load modulation using the GaAs $0.25 \mu m$ pHEMT technology to reach an enlarged output power back-off (OPBO) with circuitry solutions in order to overcome technology restrictions. To prevent power leakage in auxiliary PA ($PA_{aux}$) due to its extremely large off-state impedance, a Wilkinson power combiner is added to the output. Moreover, an input asymmetric power divider is designed to guarantee that no considerable power is delivered to main PA ($PA_{main}$) in the high-power region to make it saturated. A two-section matching network is proposed for $PA_{main}$, which simultaneously compensates for phase differences of the main and auxiliary amplification paths. To control the significant impedance variation of $PA_{aux}$ versus sweeping power and the different impedance trajectories of the main and auxiliary amplification paths, the bias and dimension selection of $PA_{aux}$ are analyzed to reach the desired output power profile versus input power. These methods overcome impedance variations and linearity degradation. To achieve the aimed 10% fractional bandwidth, appropriate low-quality LC-networks are selected as matching networks. The simulation results indicate the utility of the proposed structure for microwave link applications. Continuous-wave simulations imply that the Doherty PA has a 33dBm maximum output power and a 13.5dB power gain with less than 1dB power gain compression in the desired frequency range (7.6-8.4GHz). The drain efficiency of 30% at the highest input power, minimum of second and third harmonic powers of -140dBm and -130dBm, respectively, and OPBO of 7.5dB are also obtained.

---

[1] Corresponding author.
E-mail addresses: smrouhanie@elec.iust.ac.ir (S.M.Rouhanie), a_ghanaatian@ee.iust.ac.ir (A.Ghanaatian), abrishamifar@iust.ac.ir(A.Abrishamifar),m_tayarani@iust.ac.ir(M.Tayarani)

**Keywords**: Asymmetric, Back-off, Doherty power amplifier, MMIC, PAPR, Wideband

## 1. Introduction

To overcome the issue of efficiency roll-off in basic power amplifiers (PAs), a Doherty power amplifier (DPA) was introduced as a circuit-level solution [1]. Due to the potency of DPA for performance optimization at output power as well as its bandwidth and efficiency scopes, numerous researches have been conducted on its various structures [2-14]. One such structure is asymmetric DPA (ADPA) [2], in which $PA_{aux}$ more contributes to output power compared to $PA_{main}$. This asymmetric power ratio leads to the enlarged output power back-off and consequently the increased PA's high-PAPR signal capability. One of the critical challenges in designing DPA as a load modulation-based PA is the realization of a proper load modulation in the high-power region. In the conventional DPA, this is done with a quarter wavelength transmission line ($\lambda/4\ TL$), which has an inherently narrowband behavior. From a different aspect, before the high-power region, $PA_{aux}$ is inactive and also the transistor $Z_{off}$ is extremely small and is modeled by $C_{out}$. This impedance loading on the $PA_{main}$ output results in power leakage into $PA_{aux}$. To prevent this, an absorption technique is used, which is also a narrowband method. In the condition that $C_{out}$ is not extremely large, this method is practical. However, in the GaAs technology, due to its extremely large $C_{out}$, it is not an appropriate choice for the wideband approach. Therefore, it is mandatory to use a strategy for load modulation, by which an enlarged output power back-off like that of an asymmetric wideband ADPA is reached.

In this work, to realize a design without load modulation, the power divider is designed in a manner so that no considerable power is delivered by the main branch in a high-power level and the non-saturation status of $PA_{main}$ is guaranteed. At the output, a power combiner is used to increase the isolation of these paths, which contributes to preventing the power leakage. Meanwhile, the phase difference of the main and auxiliary amplification paths, which is conventionally compensated by adding a $\lambda/4\ TL$ to the $PA_{aux}$ input, is for the first time reimbursed in the presented output matching network of $PA_{main}$. This network is designed to present the maximum possible wideband behavior. Moreover, in terms of power boosting, this amplifier dedicates variant power amplifying regimes based on input power ($P_{in}$) level. At low-level power, $PA_{main}$ (typically operating as the class AB) has a linear amplifying regime. As $P_{in}$ increases, $PA_{aux}$ (operating in the conventional DPA as the class C) contributes to amplifying, which is called a sub-PA, hereafter. Appropriate allocation of different power gains to sub-PAs to have a linear operation is one of the key points in their design; this is more pivotal for multi-stage sub-PAs in comparison with single-stage sub-PAs. Desired linearity and appropriateness in amplifying sub-PAs depend on dimensions of transistors and their selected biases [15]. Utilizing transistors with different sizes for the wideband MMIC approach in GaAs $0.25\mu m$ pHEMT, as we did, has two challenging issues. The first issue is impedance variation with sweeping frequency and the second issue is the inconvenient variation of the transistor $Z_{out}$ when $P_{in}$ sweeps. Therefore, we discuss some extra points in the following paragraphs.

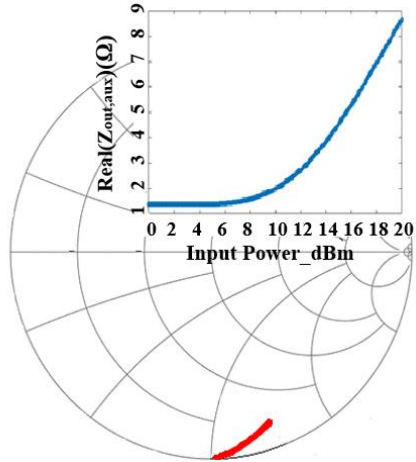

Fig.1. Second stage's $Z_{out}$ variation of auxiliary path versus sweeping power and frequency which show different trajectories.

Figure 1 indicates the output impedance trajectory of the second stage of $PA_{aux}$ versus $P_{in}$. It is interpreted that the real part of this impedance has approximately 240% proportional variation, meaning that matching networks should satisfy maximum power delivering through impedance variation in both cases of changing frequency and $P_{in}$. In this work, to overcome these limits, according to their different trends (see Figure 2), we design the first stage of the auxiliary amplifying path in terms of bias and dimension so that the best possible $Z_{out}$ trajectory can be obtained to have more matching competence with the $Z_{in}$ trajectory of the second stage of $PA_{aux}$.

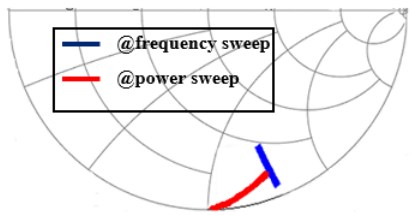

Fig.2. Second stage's $Z_{out}$ variation in auxiliary way versus sweeping input power. The realpart is shown above.

## 2. Asymmetric operation

Although designing DPA has important challenges with regard to the use of proper load modulation methods and amplifying arrangement of sub-PAs, especially in the integrated circuit approach, its suitable potential to amplify high-PAPR signals leads to the continuation of this PA inquiry and development of different approaches.

The scientific research trend of PA design and implementation indicates notable researchers' interest in introducing modified DPA structures [12-25]. Many modifications are done by changing the concept of the conventional DPA. For example, equal output powers for $PA_{main}$ and **$PA_{aux}$** are assumed in the conventional DPA, which can control the intensity of efficiency roll-off in the high-power region by allotting different output powers to them. This can be done by increasing the size ratio of sub-PAs, according to the equation (1) [26].

$$P_{Back-off} = -10.\log(1+\delta^2) \tag{1}$$

where $\delta$ is the size ratio of $PA_{aux}/PA_{main}$.

Disturbing this symmetry in both power levels using the GaAs technology with intense impedance variation when input power sweeps also leads to some other issues. Therefore, further considerations are necessary to counterbalance such issues, which are discussed in the following sections.

## 3. Design considerations

### 3.1. Topology of the circuit

The first point in MMIC design is the frequency stability of the circuit [26]. A transistor by default is an unstable element and thus using a stabilizer network is mandatory. One of the most famous stabilizer networks is a parallel RC circuit. In calculating its components, for gain roll-off prevention at higher frequencies, the capacitor of the stabilizer network should be as large as possible. However, a large capacitor for $PA_{aux}$ leads to an undesirable amplification. Moreover, it disturbs $PA_{aux}$ linearity and thus the assigned capacitor should have a relatively low capacitance, which leads to gain reduction. To make a compromise between gain reduction and linearity degradation, a two-stage amplifying topology is selected for the auxiliary path. The class AB can be designed in a single-stage structure; however, the phase difference compensation of both output signals will be highly sophisticated and is achieved with a large network, which reduces the total bandwidth. Therefore, the class AB is designed in a two-stage structure. This helps in improving overall linearity and increases circuit flexibility in order to optimize the capability of high-PAPR signals.

### 3.2. Transistor dimension assignment

As construed from the equation (1), satisfying the desired amount of main and auxiliary unequal output powers is done by setting proper dimensions of transistors according to their output power ratios. This method is realizable for single-stage sub-PAs with rational impedance trajectories in frequency and input power sweeps. However, in the case of rapid changing impedances of multistage PAs, more considerations are needed to be taken into account. To prevent the saturation of following stages, the previous stages are designed with lower dimensions than those of the second stages. At low-level $P_{in}$, the first stage of $PA_{main}$ should receive most of the power and amplify it. Since, in stabilizing an extremely short-dimension transistor, a large resistor is necessary, amplifying is not accomplished in a tolerable trend, even in the presence of a proper capacitor, and it will be acceptable only at high-level $P_{in}$. Accordingly, in this work, a smallest size transistor that can be stabilized by a rational resistor is selected ($4\times 150\ \mu m$). Figure 3 shows the large-signal load mapper simulation result and also indicates the variation of the second stage of $Z_{in,aux}$ when its $Z_{out,aux}$ changes. This implies that the amount and phase of the second stage of $Z_{in,aux}$ are highly sensitive to this change. This sensitivity increases the probability of matching disturbance at the inter-stage part. By considering this fact and according to Figure 1, to facilitate the power combining process, these two amplifying paths should be as resembling as possible. For a better impedance variation control, dimensions of the both second stages are selected similarly and their biases are chosen as similar as possible. Moreover, unequal powers received by these similar stages ensure the expected rendering of their different output powers. At first, it is necessary to find proper auxiliary second stage dimensions to provide the desired output power and amplification path of this stage, although it correlates with the selection of the $PA_{aux}$ first stage dimension. That is to say due to extreme impedance variations in the auxiliary amplification path, first, the size of the second stage sub-PA of $PA_{aux}$ will be chosen and then the size of the first stage sub-PA of $PA_{aux}$ will be selected so that the more capable $Z_{out, first\ stage}$ trajectory can be presented with the $Z_{in, second\ stage}$ trajectory. In this regard, using source and load-pull simulations with swept input power and fulfilling them repeatedly for the PA operating frequency band show that two parallel $8 \times 150\ \mu m$ transistors can present rational impedance variation in power and frequency sweep aspects. As a result, designing OMN is more realizable using a proper output power and phase trend. Given the design goal that more output power is expected from $PA_{aux}$, even if it receives more input power, because of intensive biasing, the similar dimensions of the second stages of $PA_{main}$ and $PA_{aux}$ imply that the first stage of $PA_{aux}$ will be considerably larger than the first stage of $PA_{main}$. Moreover, two stages of $PA_{aux}$ should have relatively similar dimensions to participate in the amplification process with close power gains, and the second stage is not assigned to completely perform power amplification. As shown in Figure 4, by setting the size of the first stage of $PA_{aux}$ as approximately similar to that of its second stage, impedance variations are closer in the two stages in comparison to the situation where the first stage is taken as a low-size driver. Indeed, this similarity in variations leads to similar designing of the both stages. Thus, reaching the desired impedance corresponding to the output power profile becomes more feasible. However, when the first stage is designed as a low-size driver, the power profile is only reached when the two stages have negative impedances, which is not acceptable, and their high impedance difference results in two completely different stages that barely match. Moreover, this can lead to intensifying the phase difference with the main amplifying path; therefore, the phase compensation worsens at the output. Under these circumstances and by considering the $PA_{aux}$'s swept input power and frequency source-pull results, two parallel 6×150 $\mu m$ transistors are selected to operate as the first stage of $PA_{aux}$. As a graphical investigation, Figure 5 shows the transistors size influence on the output power contours of the second stages of sub-PAs. The similarity of these stages leads to further analogy of the contours. It means that variation of adjacent paths results in a more flexible combination of their output powers. Figure 6 indicates the main and auxiliary amplifying trends according to the described conditions. Based on the desired unequal amplifying trend of sub-PAs, the absorbed input power by the main path is extremely low in spite of its amplifying capacity, and at the high-level power, almost all the power is absorbed by the auxiliary path. For the linear operation insurance of quasi-ADPA, gain compression should be traced when the nonlinear component of this structure, i.e., $PA_{aux}$, commences the amplification process. Figure 7 shows that gain compression is less than 1 dB, implying the linear amplifying of the proposed quasi-ADPA.

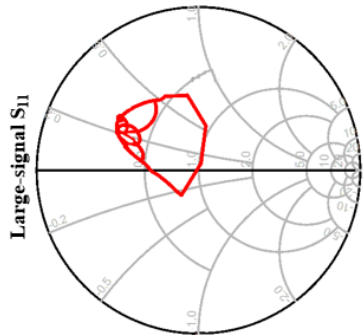

Fig.3. Influence of large-signal $S_{11}$ variation on $Z_{out}$ variation

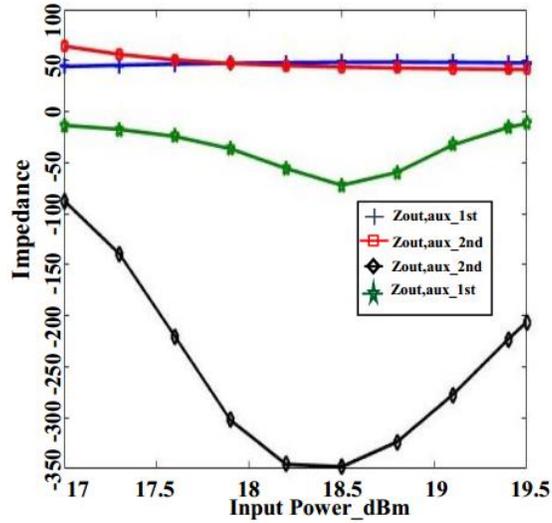

Fig.4. The $Z_{out}$ of first and second auxiliary stages. Negative impedances are related to taking the first stage as a low size driver

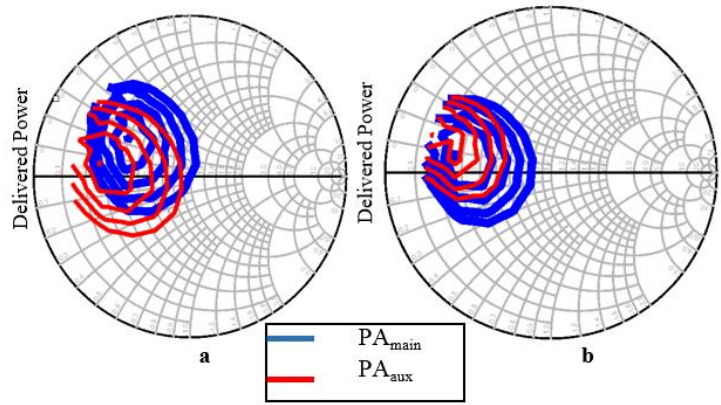

Fig.5. Delivered power contours of both sub-PAs' second stages: a) in case of $\delta = 2$  b) $\delta = 1$; more overlapped contours means more simplicity of power combining

### 3.3. Analysis of auxiliary power amplifier (PA) biasing

As previously described, the two-stage amplifying method is chosen. The next major step is to select their appropriate bias.

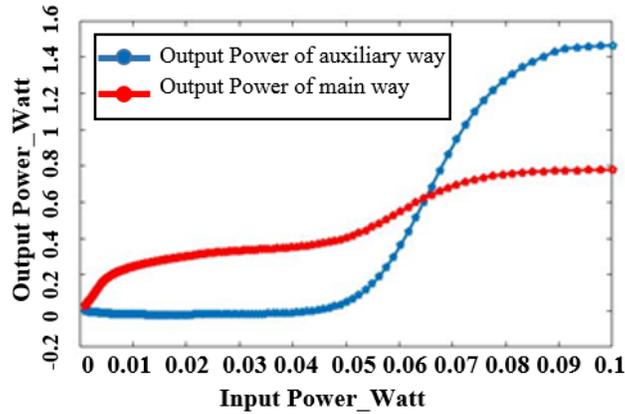

Fig.6. Output power curve of both amplifying ways

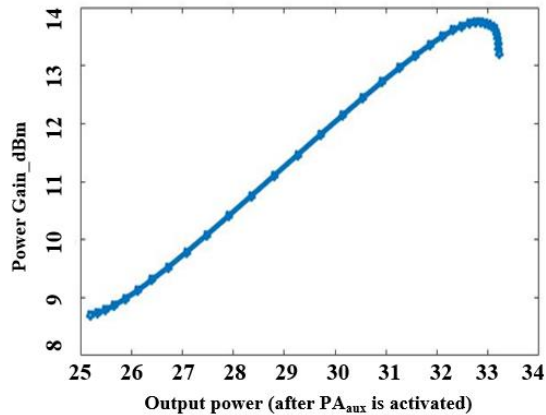

Fig.7. Power gain curve

The bias of sub-PAs should help in achieving the best possible degree of linearity, efficiency, and $P_{out}$. The main path is to receive a slight portion of $P_{in}$ and thus it is better to select the first stage amplifier in the class A mode. This selection leads to linear driving and then the second stage amplifier is operated in the class AB regime with a larger size to carry out amplifying as it is performed in a conventional DPA. There are some delicate considerations about the bias of $PA_{aux}$ beyond its conducting angle and $P_{out}$. This bias not only affects its optimum impedance for the desired $P_{out,max}$ or its correlated efficiency, but also defines the variation path of the first stage of $Z_{out}$ in the frequency or $P_{in}$ sweep. This variation affects the second stage impedance behavior and thus it is necessary to investigate $PA_{aux}$ bias selection more carefully. Three conditions are possible for this biasing approach as follows:

Case 1: The first stage is biased more intensively.

Case 2: The two stages have the same biasing scheme.

Case 3: The second stage is biased more intensively.

As described, the purpose of the two-stage structure in this work is to split the amplifying roll to avoid power roll-off in each stage due to the relatively small stabilizer network capacitor. It should be noticed that all biases deeper than the class A result in signal clipping and as a consequence lead to emerging undesired harmonics in the input signal of the amplifier. Emerged harmonics have a lower power level if this clipping occurs in the first stage, which receives a low-power signal, compared to the other conditions and thus it is compensated more easily. In the second case, amplified harmonics of the first stage along with harmonics emerged in the second stage are amplified with the second stage gain, which is larger than the first stage gain. As a result, harmonics with a higher power level emerge at the output. Finally, in the third case, since hard clipping occurs on the amplified first stage signal, harmonics with a higher power level are generated and amplified by the second stage gain and then emerge at the output.

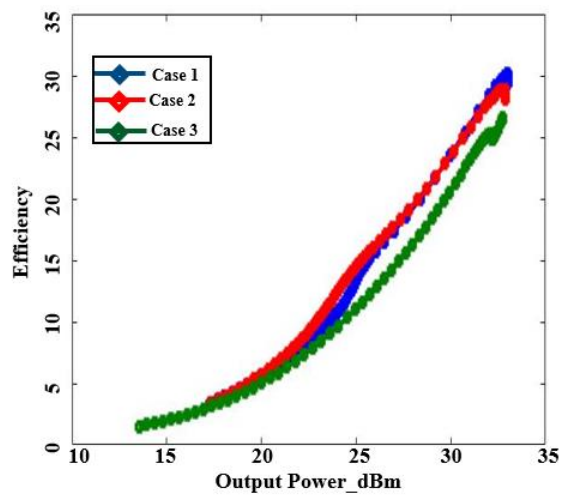

Fig.8. Drain efficiency of three described conditions for auxiliary sub-PAs biasing

One can deduce in this case that harmonics have a higher amplitude compared to the first case and thus the first case is selected for auxiliary PAs. This also affects the quasi-ADPA efficiency. As shown in Figure 8, undesired harmonics in the output power lead to a more efficient performance.

### 3.4. Input power dividing

As discussed before, due to the bandwidth degradation effects of large $C_{out,PAaux}$ absorption and the colossal power leakage into $PA_{aux}$, a power combiner is used instead of the $\lambda/4$TL. To assure the non-saturation regime for $PA_{main}$, the $Z_{characteristic}$ of the main branch of the power divider is matched to its $Z_{opt}$ in low-level $P_{in}$, and its auxiliary branch is matched to its $Z_{opt}$ in high level $P_{in}$. Therefore, when power increases, a high mismatch prevents $PA_{main}$ to receive a considerable degree of power and thus it will not become saturated according to [27]. However, in this situation, linear asymmetric power dividing faces two prominent challenges. The first challenge is a considerable difference in the $Z_{in}$ of the two sub-PAs while the second challenge is their significant variations in different impedance trajectories when receiving variable power. The both challenges lead to high inequality in the power divider elements which imposes a considerable phase difference to the both amplification paths. In this regard, phase equalization at the output will be complicated. Compensation of this high phase difference is discussed in the section 3.7. Moreover, selection of 50Ω for the characteristic impedance of the power divider branches will result in extremely low capacitances in the lumped-element model of the transmission lines of the power divider, introducing significant parasitical effects. To avoid using of these small capacitors, $Z_{characteristic}$ is set to 25Ω and the least capacitances with ignorable undesired effects (in this case, 110 *f*F) are set; then, other elements are attained corresponding to them. A low-pass filter is used for 25Ω to 50Ω matching at the input.

Finally, the low-pass filter is optimized for the maximum possible matching to the main and auxiliary input impedance trajectories by sweeping $P_{in}$. To avoid DC power penetration, DC blocks are put at the both terminals of the power divider. Since the auxiliary sub-PA power gain varies intolerably in the presence of a large capacitor, a resonance LC-network is placed to reduce its destructive effect. To have more similarity in the both amplifacation paths, the same LC-network is used for the main path (see Figure 9).

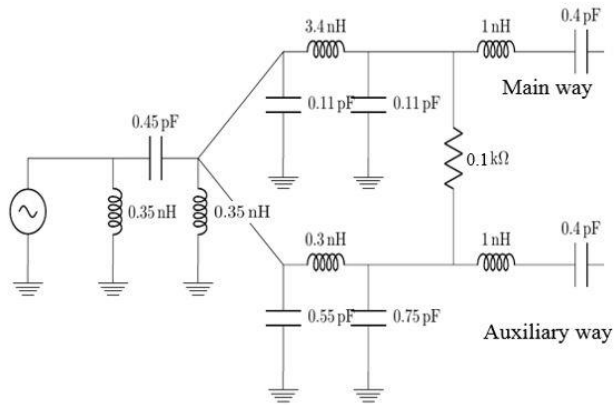

Fig.9. Schematic of input power diving with DC blocks

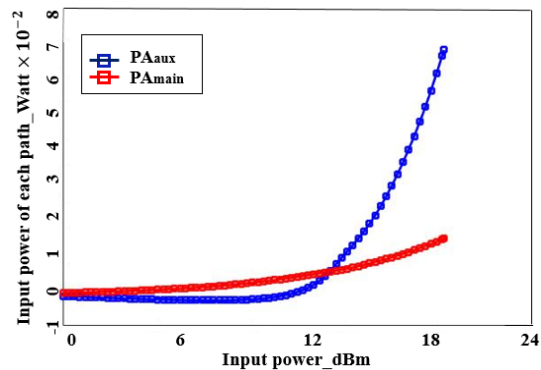

Fig.10. Delivered power to main and auxiliary ways.

The power splitting of the described power divider is shown in Figure 10. In this quasi-ADPA, all the inductors are realized by TLs due to the maximum current density. Figure 11 indicates the first stages of the proposed quasi-ADPA.

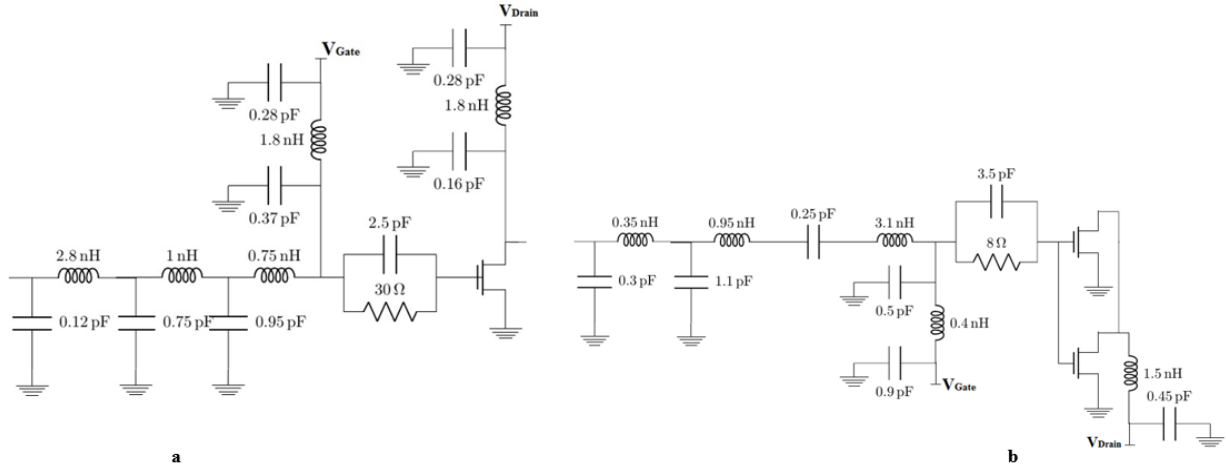

Fig.11. Schematic of first stages of both paths: a) main way, b) auxiliary way

### 3.6. Inter-stage matching networks

Like other parts of the proposed quasi-ADPA, reasonably low-quality LC-networks presenting an acceptable wideband performance are selected. To deliver the maximum $P_{out}$ of the first stage to the second stage, the optimal output impedance ($Z_{out,opt}$) of the first stage should be matched to the optimal input impedance ($Z_{in,opt}$) of the second stage, which are both elicited from load/source-pull results, respectively. However, the both impedances are complex numbers and the wideband matching of complex impedances is sophisticated. A more straightforward design strategy is to match the both impedances to a real impedance. The real impedance should be selected so that it can match $Z_{out,opt}$ and $Z_{in,opt}$ with fewer elements in a low-quality path in the Smith chart. For the both amplifying paths, these impedances are selected to be 50Ω. These matching networks are depicted in Figure 12.

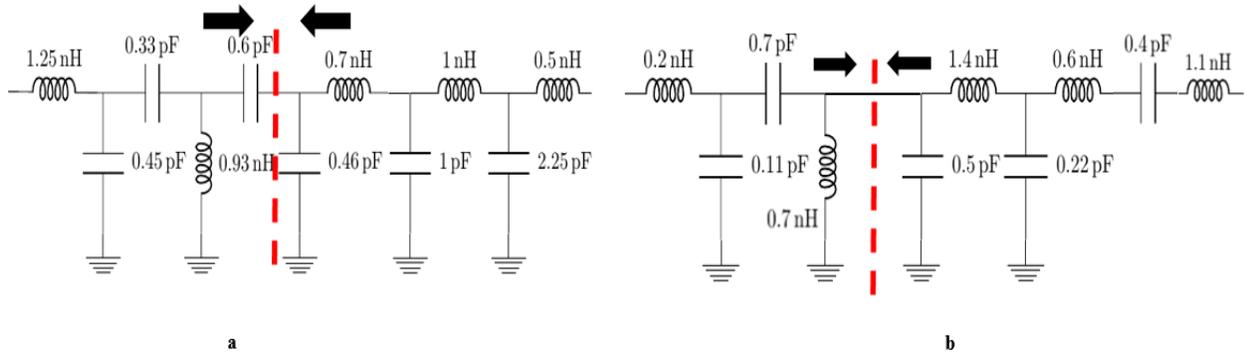

Fig.12. Schematic of inter-stages of both paths: a) main way, b) auxiliary way

### 3.7. The compensating phase and matching network selection

The conventional DPA concept is based upon a hypothesis that when $PA_{aux}$ is off, its output impedance tends to be infinite and no power leakage occurs from $PA_{main}$ into it. However, it will be necessary to find a proper technique to increase the output isolation if off-state $Z_{out,aux}$ is not sufficiently large in the real world and absorbing techniques are not practical due to the desired bandwidth. For combining the maximum power, in this work, two conditions should be satisfied as follows:

1. Adequate isolation of terminals of the both amplifying paths to avoid power leakage
2. Acceptable wideband performance

By considering the trajectory of $Z_{out,aux}$, the $\lambda/4$ TL meets none of them. Therefore, it is imperative to use a wideband matching network by adding an extra block to satisfy adequate isolation, which will be discussed in the next section. To have an efficient structure, the delivered output powers of the both amplification paths should be in-phase or at least have a tolerable phase inequality. Since here, its value is not as low as possible to be ignored (almost 120 degrees), phase and impedance matching equations should be considered simultaneously. In other words, in this case for the first time, the phase compensating and impedance matching networks are merged at the end of these two paths. There are three approaches to design the phase compensating network as follows:

Case 1: Merge in the OMN of the class AB

Case 2: Merge in the OMN of the class C

Case 3: Merge in the power combiner

The second case is not appropriate since it intensifies the inherent narrowband characteristic of the class C PA. In other words, it results in quasi-ADPA bandwidth reduction, and its unconventional impedance trajectory makes it highly sophisticated to combine its equations with that of the phase compensation. In the third case, linear power combining has poor controllability on phase compensation and merging these two networks will result in sacrificing the linearity. Moreover, proper power combining will occur only at high output power level. Thus in the first case, to make a trade-off between the increased wideband performance of this OMN and its decreased insertion loss, a two-section matching network is selected. The aimed value of the phase that should be compensated is split between them, which are modeled by two TLs with the parametric $Z_{characteristic}$ and electrical lengths. To this end, $Z_{out,main}$, 10.6+j5.7, should be matched to a 50Ω resistor using a two-step impedance transfer function. To do so, 10.6+j5.7 is first matched to a 25 Ω resistor and then to a 50 Ω resistor. In other words, these close impedance ratios of 2.3 and 2 are selected to present similar performances in terms of bandwidths. As a result, the optimization of the network will be facilitated. After that, impedance matching formulas are written as an objective function. The desired phase compensation is considered as a constraint equation for the objective function. Finally, a constrained genetic algorithm is applied using MATLAB to obtain results. The outputs of the genetic algorithm are two characteristic impedances and two electrical lengths. Based on the transmission line theory, the following equations lead to finding the object function. For the first TL, we have:

$$10.6 + j5.7 = Z_{o1} \frac{25 + jZ_{o1} \tan(\theta_{o1})}{Z_{o1} + j25 \tan(\theta_{o1})} \tag{2}$$

Then, for the second one, we have:

$$25 = Z_{o2} \frac{50 + jZ_{o2} \tan(\theta_2)}{Z_2 + j50 \tan(\theta_2)} \tag{3}$$

Therefore:

$$\text{Object Function} = 25Z_{o1} + j(Z_{o1}^2)\tan(\theta_1) - (25Z_{o1} + j625\tan(\theta_1)) + 50Z_{o2} + j(Z_{o2}^2)\tan(\theta_2) - (50Z_{o1} + j2500\tan(\theta_2)) \tag{4}$$

In the circuitry model of the transmission line, to achieve realizable elements (especially appropriate capacitors) in this technology, a low-pass filter is chosen with series inductors. Figures 13 and 14 indicate the schematics of this network and the output stages of the both paths, respectively.

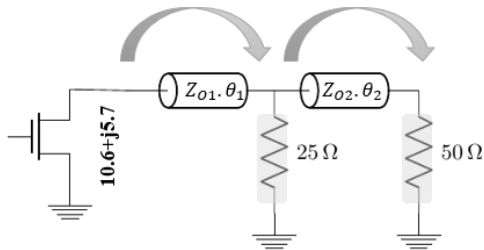

Fig.13. Proposed OMN schematic (resistors are put to show target impedance of each section)

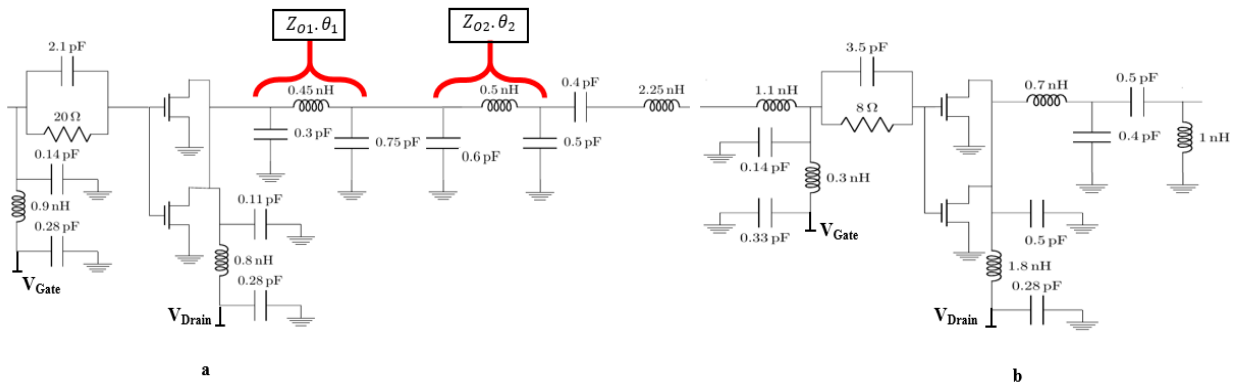

a

b

Fig.14. Schematic of second stages of both ways: a) main path, b) auxiliary path

### 3.8. Output power combining

As described previously, unlike the DPA theory where a $\lambda/4$TL is used to carry out the desired load modulation, by neglecting the $\lambda/4$TL narrowband behavior, on the condition that transistor output impedances are not sufficiently high to make sure that no power leakage occurs, designers are compelled to change the power combining structure.

One of the appropriate structures is the inverted DPA structure [14]; however, this solution suffers from the narrowband behavior of the transmission line. Furthermore, the circuitry model of the transmission line in a high-center frequency leads to the use of extremely large capacitors, which are not appropriate for the MMIC design. Therefore, in this work, the output powers of the amplifying paths are combined by using a power combiner with the proper isolation of its terminals. Because of the acceptable isolation and wideband behavior of the Wilkinson power combiner, an asymmetric Wilkinson combiner is chosen to combine the main and auxiliary output powers with a power ratio of about $0.5\bar{3}(P_{out.main}/P_{out.aux})$. The schematic of the asymmetric combiner is shown in Figure 15. This ratio leads to a nuance in the phase of each power combiner terminal.

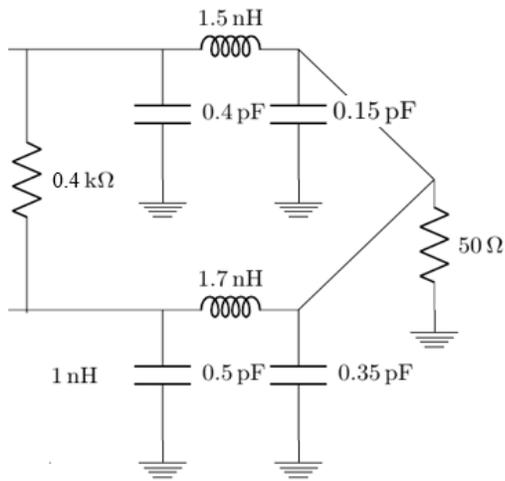 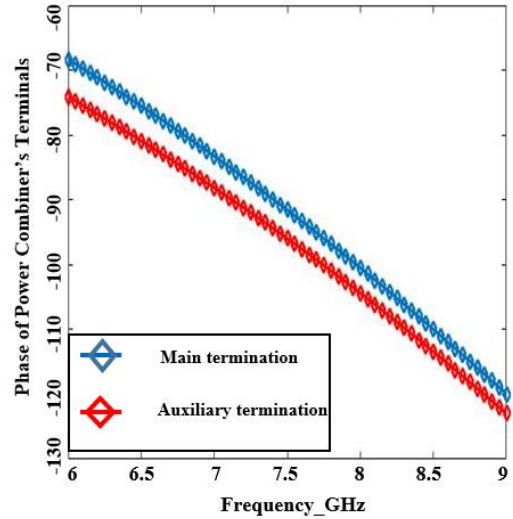

Fig.15. Output power combiner schematic          Fig.16. Phase of power combiners' terminations

It can be interpreted from Figure 16 that this difference is reasonably small and thus there is no concern about unsatisfying power combining. By using this approach, Figure 17 depicts the proposed quasi-ADPA drain efficiency, output power, and power gain. Moreover, the drain efficiency of the proposed quasi-ADPA at different frequencies is indicated in Figure 18.

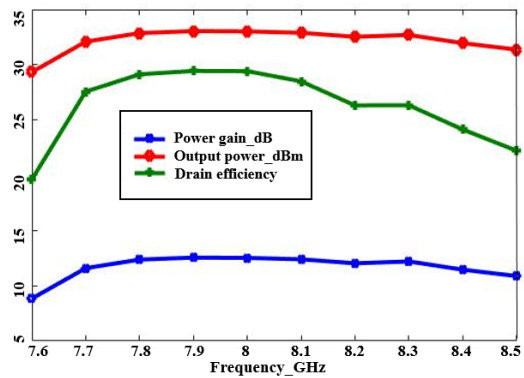

Fig.17. Output power, drain efficiency and power gain of DPA

### 3.9. The study of the second and third harmonics

Power leakage to undesirable harmonics by a harmonic cancelation network, regardless of its causes, should be prevented. According to [15], based on the network effect on the bandwidth, it is better to put this network at the main amplifying path. This will increase the phase difference of the paths. In this work, power in the second and third harmonics is too low so that the harmonic cancelation network design can be ignored. This is checked by imposing a continuous-wave-signal, which shows that almost no amplifying occurs at the second and third harmonics. As a result, the harmonic cancelation network design is skipped, as shown in Figure 19.

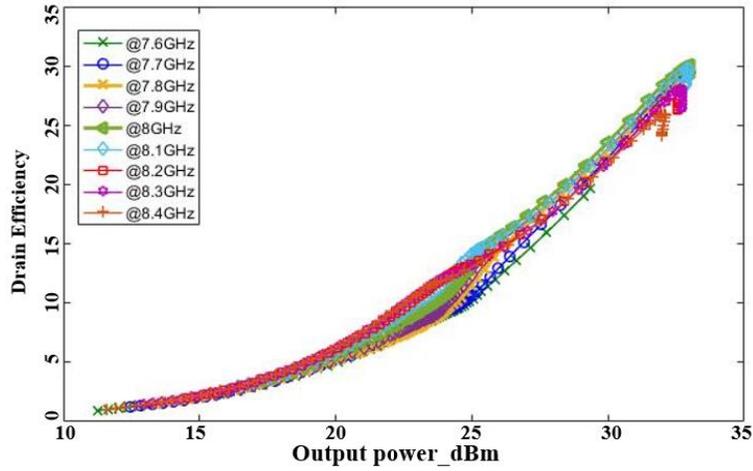

Fig.16. Drain efficiency curve at desired frequencies

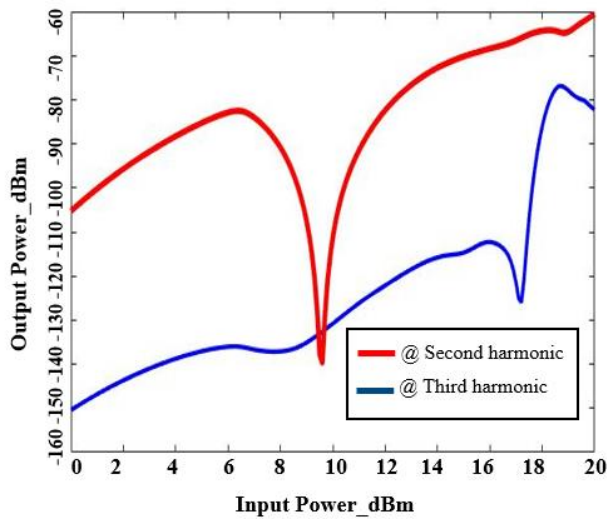

Fig.19. Output power in second and third harmonics

## 4. Results

According to the ADPA theory, to design a PA with increased output power back-off, an quasi-ADPA is designed with two main and auxiliary paths using the GaAs $0.25\mu m$ pHEMT monolithic technology. These amplifying paths present maximum power gain of 13.5 dB, which leads to $P_{out,max}$ of 33 dBm and drain efficiency of 30% for the maximum input power when the main path drains biases are 4 V and 5.5V for its first and second stages and the auxiliary path drains biases are 5V and 7.25V for its first and second stages, respectively. The dimensions of these sub-PAs are selected $4 \times 150 \ \mu m$, $8 \times 150 \ \mu m$, $6 \times 150 \ \mu m$ and $8 \times 150 \ \mu m$ for the first and second stages of the main and auxiliary amplifying paths, respectively, according to the best results of their swept input power source/load-pull simulations. To achieve proper circuitry elements with the lowest parasitic and destructive influence on bandwidth and linearity, a two-section LC-matching network with a sufficiently low quality is proposed to match and compensate for phase differences in the both paths. To prevent output power leakage into the auxiliary path due to its low off-state impedance, a Wilkinson power combiner is used. As a result, the desired 10% fractional bandwidth is attained at the 7.6-8.4GHz frequency range. The gain compression is less than 1 dB, which implies the linear operation of PA. The power study of the minimum second and third harmonics demonstrates that there is no significant power in these harmonics, -140dBm and -130dBm, respectively, and thus almost all the output power is in the desired harmonic. The total layout of quasi-ADPA is indicated in Figure 20. The performance of the proposed quasi-ADPA is compared with the other recently reported MMIC ADPAs in Table 1.

Table.1. Comparing the designed quasi-ADPA performance with other recent MMIC ADPAs

|  | [9]** | [10]*** | [11] | This work |
|---|---|---|---|---|
| Technology | InGaP/GaAs HBT | GaAs pHEMT0.15um | GaAs pHEMT0.15um | GaAs pHEMT0.25um |
| Frequency(GHz) | 0.2-1.5 | 25.8 | 24 | 7.6-8.4 |
| Peak Power(dBm) | 25 | 25.1 | 32.78 | 33 |
| **OPBO(dB)** | < 6 | < 1 | 6 | **7.5** |
| Peak Drain Efficiency(%) | 20 | > 16**** | 35 | 30 |
| Power Gain(dB) | 30 | 7 | 11.5 | 13.5 |

*Simulation results are reported
** For Power Gain less than 3 dB
*** Measured results, simulation ones are not reported
**** It is estimated

## 5. Conclusion

In this work, for the first time, a quasi-ADPA was designed without load modulation. This PA was designed using the GaAs $0.25 \mu m$ pHEMT technology, which confronted serious technological bottlenecks. All efforts were based on a step by step circuit-level solution to investigate and overcome these limitations. The size and power delivering rate of transistors to each path were arranged to attain a 7.5 dB output power back-off. According to [27], the asymmetric Wilkinson power divider was used in the input to match the PA$_{main}$ branch to Z$_{in,main}$ at low-level input power when the auxiliary branch was matched to Z$_{in,aux}$ at high-level input power. Thus in low-level power, the auxiliary branch was mismatched to have a proper amplifying trend later at high-level power, and when P$_{in}$ reached its maximum amount, the mismatch was minimum. Then, a multistage structure with the coarse to fine biasing of auxiliary PAs prevented linearity and efficiency degradation due to the non-existence of large capacitors in the stabilizing network and also prevented emergence of undesired harmonics. Finally, our proposed matching network, which was inserted at the end of the main amplifying path, simultaneously compensated for the phase difference of these paths. Moreover, a Wilkinson power combiner blocked power leakage into the auxiliary amplifying path when it was inactive. By these strategies, the designed wideband quasi-

ADPA was simulated. The result of EM simulations of the proposed quasi-ADPA showed the maximum output power of 33 dBm, drain efficiency of 30% and power gain of 13.5 dB on the condition that power gain compression was less than 1dB. As a result, the quasi-ADPA linear performance was insured. Further, the minimum second and third output power harmonics of -140 dBm and -130 dBm implied that no destructive harmonic was in the output power.

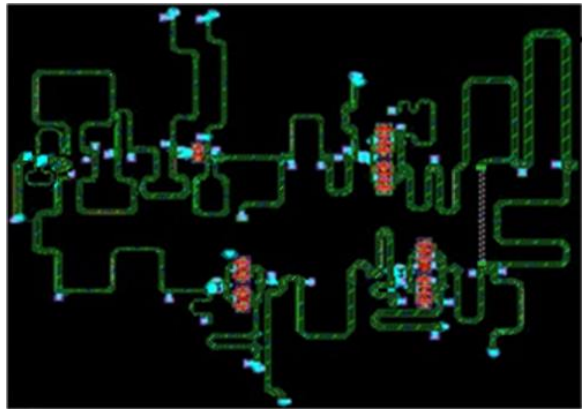

Fig.20. Total PA layout

**Acknowledgment**

The authors want to appreciate Baset Mesgari for the priceless guides on doing EM simulations.